**Diffractive imaging of a molecular rotational wavepacket with femtosecond Megaelectronvolt electron pulses**

Jie Yang[1], Markus Guehr[2], Theodore Vecchione[2], Matthew S. Robinson[1], Renkai Li[2], Nick Hartmann[2], Xiaozhe Shen[2], Ryan Coffee[2], Jeff Corbett[2], Alan Fry[2], Kelly Gaffney[2], Tais Gorkhover[2], Carsten Hast[2], Keith Jobe[2], Igor Makasyuk[2], Alexander Reid[2], Joseph Robinson[2], Sharon Vetter[2], Fenglin Wang[2], Stephen Weathersby[2], Charles Yoneda[2], Martin Centurion[1], Xijie Wang[2]

[1]University of Nebraska-Lincoln, 855 N 16th Street, Lincoln, Nebraska 68588, USA

[2]SLAC National Accelerator Laboratory, Menlo Park, CA 94025, USA

*Address correspondence to:

jie.yang@huskers.unl.edu

mguehr@stanford.edu

martin.centurion@unl.edu

wangxj@slac.stanford.edu

Imaging changes in molecular geometries on their natural femtosecond timescale with sub-Angstrom spatial precision is one of the critical challenges in the chemical sciences, since the nuclear geometry changes determine the molecular reactivity. For photoexcited molecules, the nuclear dynamics determine the photoenergy conversion path and efficiency. We performed a gas-phase electron diffraction experiment using Megaelectronvolt (MeV) electrons, where we captured the rotational wavepacket dynamics of nonadiabatically laser-aligned nitrogen molecules. We achieved an unprecedented combination of 100 fs root-mean-squared (RMS) temporal resolution and sub-Angstrom (0.76 Å) spatial resolution that makes it possible to resolve the position of the nuclei within the molecule. In addition, the diffraction patterns reveal the angular distribution of the molecules, which changes from prolate (aligned) to oblate (anti-aligned) in 300 fs. Our results demonstrate a significant and promising step towards making atomically resolved movies of molecular reactions.

The dynamical behavior of molecules is governed by the complex interplay of a correlated system of nuclei and electrons which interact via Coulomb and exchange forces. Determining how the individual nuclei within a molecule move relative to one another during a molecular transformation represents a key step to understanding chemical reactivity. Developments in the field of femtochemistry enabled capturing the motion of nuclear wavepackets using purely spectroscopic measurements with femtosecond laser pulses[1–11]. For small molecules, structural information on the nuclear geometry can be indirectly inferred based on exact knowledge of spectral transitions involved in the probe process. For larger molecules, this inference becomes unfeasible. Diffraction techniques, which provide direct access to the position of each atom within a molecule, are a more effective approach. Over the last few decades great strides have been taken to observe ultrafast dynamics with ultrashort x-ray pulses from synchrotrons and free electron lasers (FELs)[12–15] and via diffraction of short electron pulses[16–19]. With the advent of FELs, X-ray diffraction experiments have now reached a sub-100 fs temporal resolution. The spatial resolution is sufficient to observe larger scale molecular processes like light-induced opening of a 6-membered ring, but not to resolve individual atoms due to the limited wavelength available in FELs[15]. Laser-induced electron diffraction[20,21] and photoelectron holography[22] can also provide spatial and temporal resolution simultaneously, but require significant theoretical input to retrieve the interatomic distances. Ultrafast electron diffraction (UED) from gas-phase molecules has achieved sub-Angstrom spatial resolution but the temporal resolution has not been sufficient to observe molecular geometry changes on the femtosecond timescale[23,24]. Here we report a crucial advance in gas-phase UED using multi MeV relativistic electron pulses to achieve a temporal resolution of 100 fs RMS (230 fs full-width at half maximum(FWHM)) that opens the door to observing the motions of individual nuclei that result from structural changes during photochemical reactions of isolated molecules.

The majority of UED experiments have been performed on solid-state samples[25,26], which are ideal to understand collective effects in condensed media such as superconductivity, heat transport and magnetism. Gaseous molecules are ideal to study prototypical processes in chemistry. They also provide a direct link between experiments and quantum chemical calculations, which can be performed on the highest level for isolated molecules[26–28]. Early gas phase UED studies employed stroboscopic electron diffraction[29]. A breakthrough was achieved when picosecond electron beams became available in late 1980s[30]. Zewail and co-workers were able to resolve non-equilibrium molecular structures[31], transient molecular structures[18], and radiationless dark structures[19] using picosecond UED. However, in order to resolve molecular geometry changes in real time, a better temporal resolution is required. In the context of the crucial excited state photoisomerization reactions[27], a 200 fs resolution is suitable for exploring the nuclear dynamics of the isolated azobenzene isomerization[32]. In addition, this temporal resolution is sufficient to explore photoprotection processes of nucleobases[33–36].

To improve UED temporal resolution, two major effects must be overcome: space-charge repulsion between electrons[37], and velocity mismatch[38] resulting from the electron pulse lagging behind light pulses used for the molecular excitation. Both of these limitations can be minimized using relativistic MeV electrons[39–48]. The longitudinal space-charge pulse elongation is proportional to $1/\beta^2\gamma^5$, where $\beta=v/c$, $\gamma=(1-\beta^2)^{-1/2}$ is the Lorentz factor, and $v$ and $c$ are the speed of electrons and the speed of light in vacuum, respectively[49]. Concerning the velocity mismatch, electrons with 3.7 MeV kinetic energy travels at $v$=0.993$c$. This results in only 5 fs delay with respect to an optical pulse for a typical 200 μm interaction length.

Here we present the ultrafast laser-induced rotational dynamics of $N_2$ molecules in the gas phase. We impulsively excited a rotational wavepacket in a $N_2$ gas sample using a 34 fs FWHM laser pulse, spectrally centered at 800 nm, which interacts non-resonantly with the anisotropic molecular polarizability tensor. The phase evolution of the rotational wavepacket results in rotational revivals, with molecules alternating between aligned and anti-aligned directions[50,51]. Previous efforts to investigate rotational wavepackets with UED have captured dynamics on picosecond time scales[52,53]. The rotational dynamics of laser-aligned $N_2$ has been previously observed with optical birefringence[54], strong field ionization[55],high harmonic generation[56,57], and Auger electron spectroscopy[58]. We have observed the temporal evolution of the full wavepacket revival with an 8.35 ps period by quantifying the anisotropy in the diffraction patterns. The temporal resolution was determined by a fitting routine using the measured dynamics. Furthermore, we have retrieved molecular images of the aligned and anti-aligned molecular ensemble with atomic resolution.

## Results

**Experimental layout and static diffraction.** The experimental layout is shown in Fig. 1. The electron pulse (blue) is diffracted from the nitrogen gas jet (gray), which is introduced into the vacuum chamber using a pulsed nozzle (black).The diffraction pattern is captured by a phosphor screen and a detector. The 800nm alignment laser pulse (red) is directed to the target and removed from the vacuum chamber by two holey mirrors at a 45 degree angle to the electron beam. The full set-up is discussed in detail elsewhere[59]. More details of the experiment can be found in the Experimental Setup section in Methods.

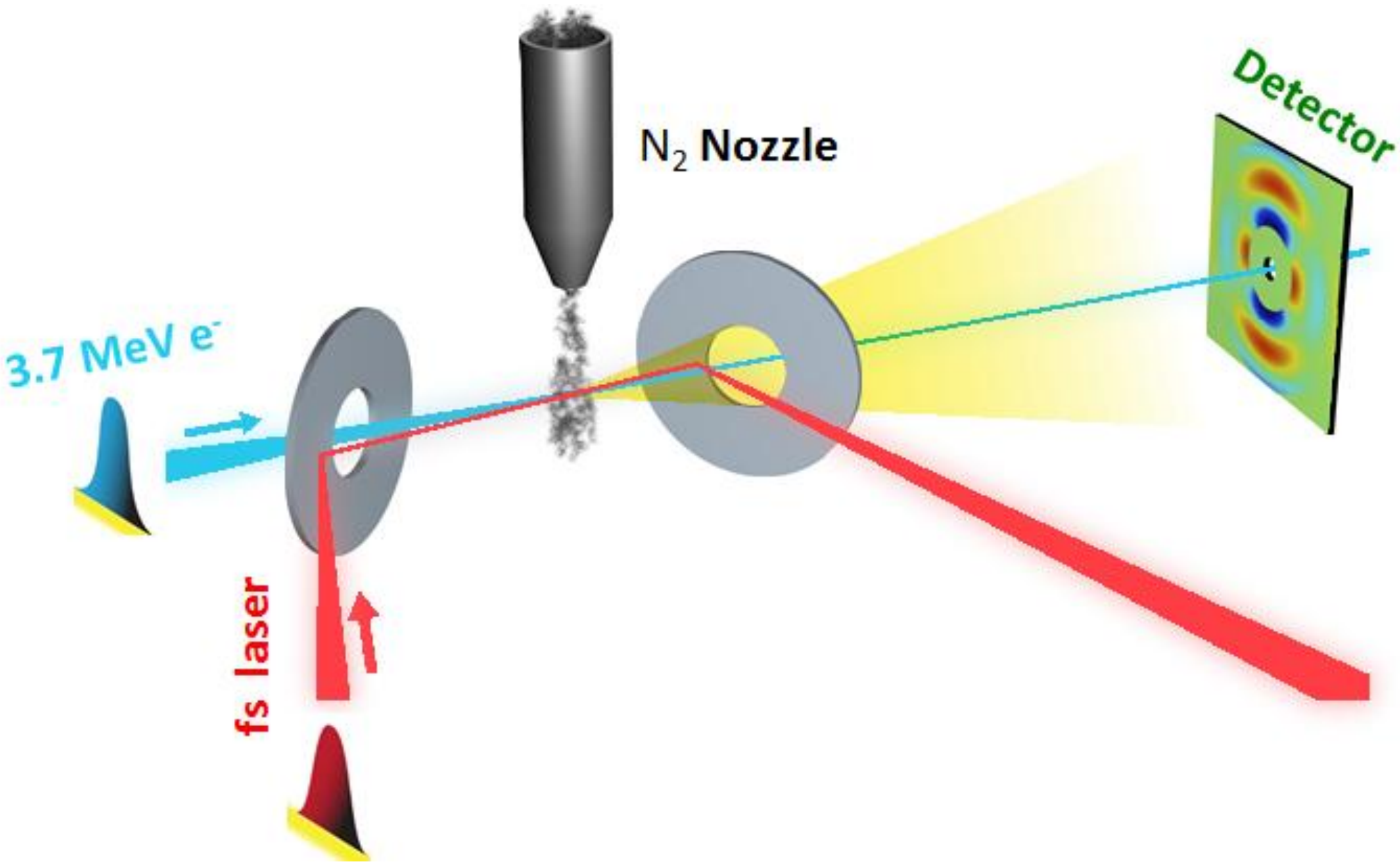

**Figure 1 : Experimental layout.**

The diffraction pattern is commonly expressed as a function of the momentum transfer

$$s = \frac{4\pi}{\lambda}\sin(\theta/2) \qquad (1)$$

where $\lambda$ is the wavelength of the electron beam, and $\theta$ is the angle between the scattered and transmitted electrons. For a 3.7 MeV electron beam, $\lambda$ =0.30 pm. The total scattering intensity $I_{tot}$ is the sum of the atomic scattering intensity $I_{at}$ and the molecular scattering intensity $I_{mol}$. $I_{at}$ is defined as

$$I_{at} = \sum_{i=1}^{N} |f_i(s)|^2 \qquad (2)$$

where $N$ is the number of atoms in the molecule and $f_i$ is the elastic scattering amplitude for the $i^{th}$ atom. For MeV electrons, $f_i$ can be calculated using the ELSEPA program[60].

The structural information of the molecule is encoded in the molecular scattering intensity, $I_{mol}$, given by

$$I_{mol} = \sum_{i=1}^{N} \sum_{j \neq i}^{N} |f_i(s)| |f_j(s)| \cos(\eta_i - \eta_j) \frac{\sin(s r_{ij})}{r_{ij}} \quad (3)$$

where $\eta_i$ is the scattering phase of the $i^{th}$ atom and $r_{ij}$ is the distance between the $i^{th}$ and the $j^{th}$ atoms[23].

The so-called modified diffraction intensity is defined as

$$sM(s) = s \frac{I_{mol}(s)}{I_{at}(s)} \quad (4)$$

The spatial resolution, $\delta$, is determined by the maximum measured $s$ value $s_{max}$ using the formula

$$\delta = 2\pi / s_{max} \quad (5)$$

Figure 2 shows the experimental and theoretical modified diffraction intensity for static diffraction from *unaligned* $N_2$. The experimental background was removed using a standard fitting procedure[23]. The bond length was determined to be 1.073 ± 0.027 Å, in agreement with the previously measured $N_2$ bond length of 1.098 Å[61]. The 2.5% measurement uncertainty is due to the uncertainty in the calibration of the sample-to-detector distance and electron energy, as explained in the Experimental Set-up section in Methods. In this experiment, scattering signal is available in the region between 3.5 and 12 $Å^{-1}$. The fitting procedure relies on the zeros of $sM(s)$ and this reduces the available data in Fig. 2 to $s > 4.5$ $Å^{-1}$. The measurement agrees well with the simulation up to s~12 Å$^{-1}$.

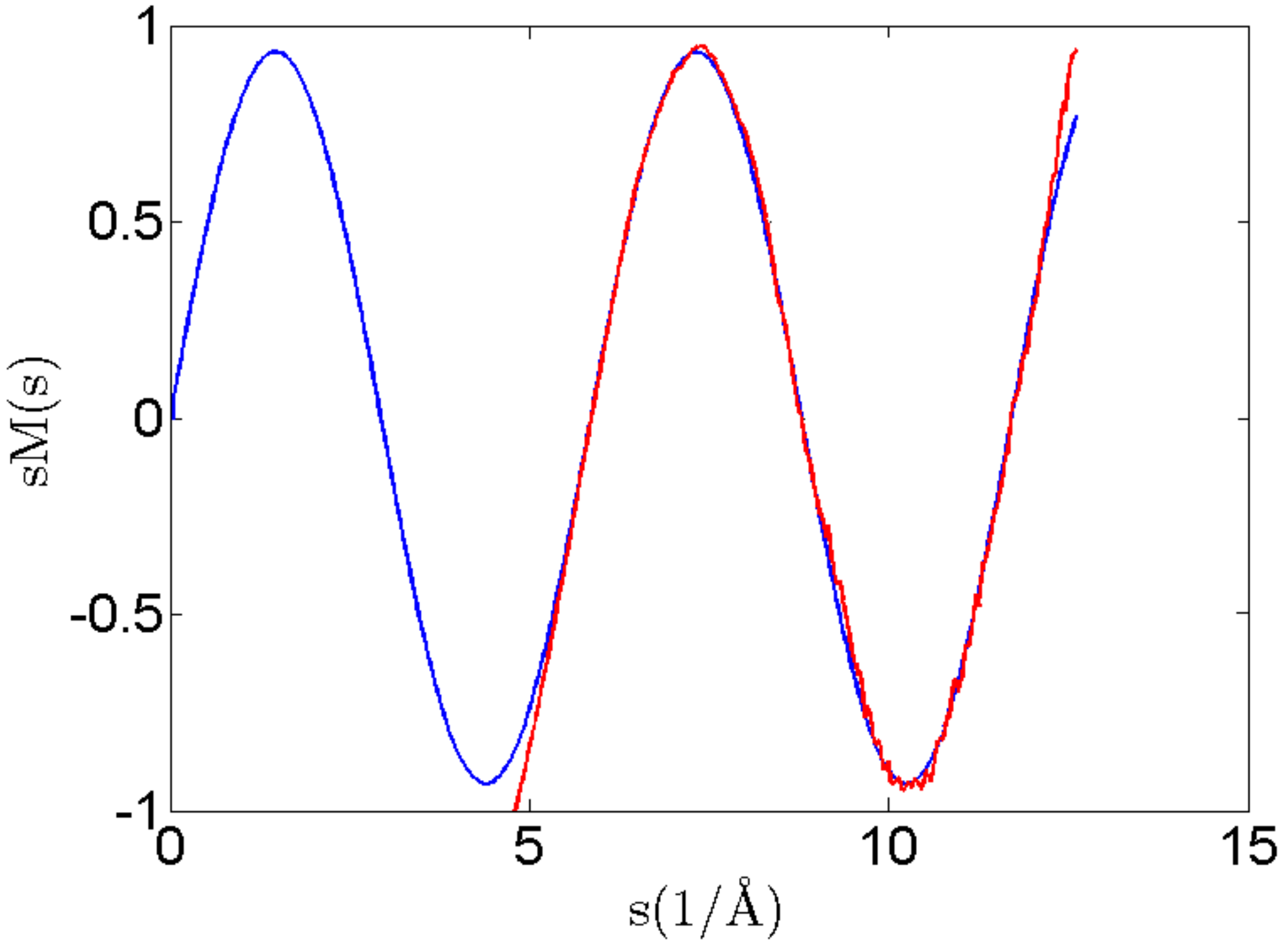

**Figure 2: Static $N_2$ diffraction.**

**Temporal evolution of $N_2$ alignment.** For impulsive alignment, the full rotational revival is expected at $t$=1/2$cB$, where $B$ is the rotational constant and $c$ is the speed of light in vacuum. For $N_2$ molecules, the full revival is at 8.35 ps ($B$=1.998 cm$^{-1}$). The rapid evolution of the angular distribution can be used to determine the temporal resolution of the measurement technique.

Diffraction patterns from a molecular ensemble aligned with a polarization in the detector plane are not circularly symmetric, contrary to the static case in equation 3. The anisotropy, $a(t)$, in a diffraction pattern can be used to trace the temporal evolution of alignment[52]. In addition, $a(t)$ is a self-normalized parameter that is extracted directly from the diffraction patterns (see Data Processing of Diffraction Patterns and Anisotropy in Methods). Figure 3 shows the temporal evolution of the simulated and experimentally measured $a(t)$ values. The simulation is composed of two parts: an impulsive alignment simulation that calculates the angular distribution at different delay times[62], followed by a diffraction pattern simulation based on the modeled angular distribution[63]. The anisotropy of the simulated patterns is calculated using the same method as for the experimental patterns. The details of the simulation are explained in the Alignment Simulation and Diffraction Pattern Simulation section in Methods. The experimental data is recorded with 100 fs steps in the delay between laser and electron pulses, and at each point data are collected for 2 minutes (14,400 shots at a repetition rate of 120 Hz).

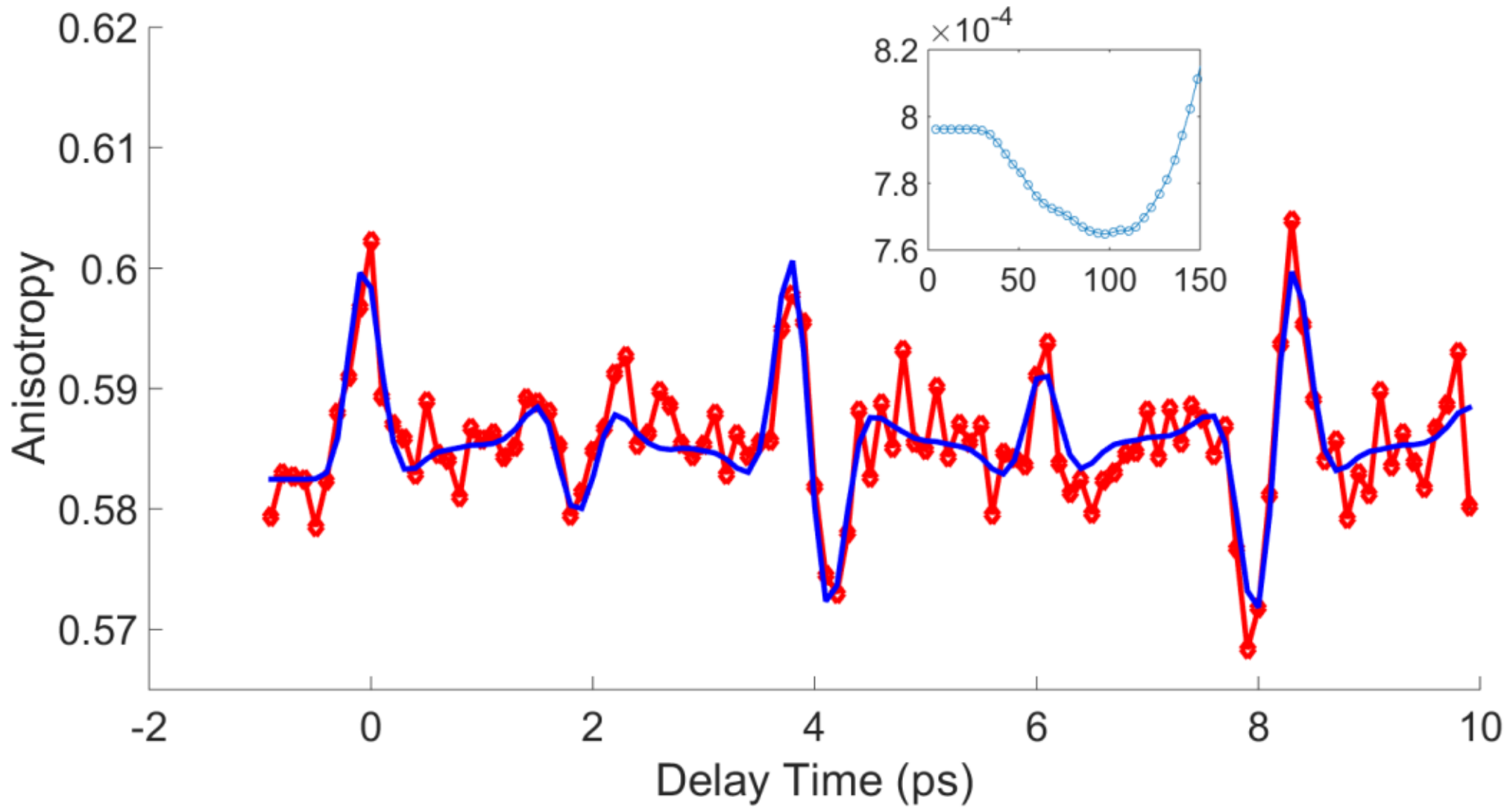

**Figure 3: Temporal evolution of the $N_2$ rotational wavepacket.**

The temporal evolution of the alignment has a rich structure that varies on a fast time scale. For example, at the half-revival (delay of 4 ps) the distribution changes from aligned to anti-aligned in 300 fs. The alignment peak corresponds to a prolate angular distribution, with the long axis along the direction of the laser polarization. The angular distribution during anti-alignment is oblate, with the molecules preferentially lying in a plane perpendicular to the laser polarization. The full revival at around 8 ps shows a similarly fast transition from oblate to prolate distribution. In between the revivals, the anisotropy of the diffraction pattern captures additional dynamics.

We have used the fast-changing distribution to characterize the temporal resolution of the measurement. The shape of the aligned molecular ensemble is determined by the initial temperature of the molecules and the fluence of the alignment laser pulse[64]. In this experiment, the laser fluence was measured to be 2.0 J/cm$^2$ and the initial temperature was estimated to be 54 K, using a supersonic expansion model[65]. At these parameters, the wavepacket revivals are relatively sharp compared to the temporal resolution. The limited temporal resolution effectively blurs out the rotational dynamics and has a significant effect on the observed structure. We performed first a two-parameter $\chi^2$ fitting by fixing the laser fluence and initial rotational temperature to the measured and calculated values, and then a full four-parameter fit where all parameters were allowed to vary.

The two-parameter fit, varying only the temporal resolution and a re-scaling factor that accounts for the spatial overlap between laser and electron pulse, returned a temporal resolution of 85 fs RMS (200 fs FWHM). The four-parameter fit, which varies temporal resolution in addition to initial rotational temperature, laser fluence, and the re-scaling factor, accounts for uncertainties in the laser fluence and initial temperature. This method achieved a best fit resolution of 100 fs RMS (230 fs FWHM), with an initial temperature of 65K and a laser fluence of 1.8 J/cm$^2$, which are comparable to our initial estimates for the two-parameter fit. The simulation shown in Fig. 3 shows the best fit for all four parameters. The fitting $\chi^2$ error in the four-parameter fit is shown in Fig. 3 inset as a function of the overall RMS temporal resolution. More details of the fitting are explained in the Temporal Evolution Fitting section in Methods.

The 100 fs (RMS) overall temporal resolution of this experiment is consistent with our expectations, based on the performance study of this machine[59]. A simulation showed that the electron bunch length was 70 fs RMS at the interaction region. Measurements of the phase and amplitude stability of the RF gun lead to an expected time of arrival jitter of 50 fs RMS[59]. The calculated overall temporal resolution was then 87 fs RMS, or 205 fs FWHM, which is close to the measured value.

**Molecular images with different angular distribution.** High resolution molecular diffraction images were retrieved for prolate and oblate ensembles at the half revival. Diffraction patterns with adequate signal-to-noise ratio were recorded with 60 and 90 minutes of integration time for oblate and prolate distributions, respectively. We use diffraction-difference patterns to remove the experimental background and the diffraction signal from unexcited molecules. The diffraction intensity difference is given by $\Delta I(t) = I(t) - I(\mathrm{t} = -5\ \mathrm{ps})$, where $t$=0 corresponds to the maximum of the first alignment peak after laser. Before $t$ = -0.4 ps when the pump laser arrives, the angular distribution is isotropic. In the 2-D diffraction patterns, the $s_{max}$ = 8.3 Å$^{-1}$, corresponding to a spatial resolution of 0.76 Å. This makes it possible to observe molecular structures with resolution better than the shortest possible bond lengths.

Figure 4 shows the experimental (left panels) and simulated (right panels) 2-D diffraction patterns and their corresponding Fourier transforms for prolate and oblate distribution at the half revival. Figures 4a and 4b show the experimental and simulated diffraction-difference patterns $\Delta I$ for the prolate distribution. The diffraction pattern was captured at a time delay of 3.8 ps after the first alignment peak (Fig. 3a). The diffraction-difference pattern is anisotropic as a result of molecular alignment. The experimental pattern shows excellent agreement with the simulation for a range of s=3.5 to 8.3 Å$^{-1}$.

Figures 4c and 4d show the Fourier transforms of the difference signals in Fig. 4a and 4b, respectively. The Fourier transform of the diffraction pattern displays the autocorrelation of the molecular structure, convolved with the angular distribution and projected onto the detector plane. The center of the Fourier transforms goes to zero because they are generated from the difference of two diffraction patterns. For a diatomic molecule, the autocorrelation is directly related to the molecular image. For more complex molecules, an image of the structure can be reconstructed using phase retrieval algorithms[63]. In the autocorrelation functions depicted in Fig. 4c and 4d, the positive regions indicate an increase in population, and the negative regions indicate a decrease. Specifically, Fig. 4c and 4d indicate that the population of molecules that are lying perpendicular to the laser polarization has decreased while the population parallel to the polarization (vertical in Fig. 4) has increased, i.e., more molecules are aligned along the vertical direction. Similarly, Fig. 4e and 4f show the measured and simulated diffraction pattern for the oblate distribution at half revival, corresponding to a time delay of 4.1 ps. Figures 4g and 4h are the Fourier transforms of Fig. 4e and 4f, which show that the molecular ensemble is aligned in the horizontal plane.

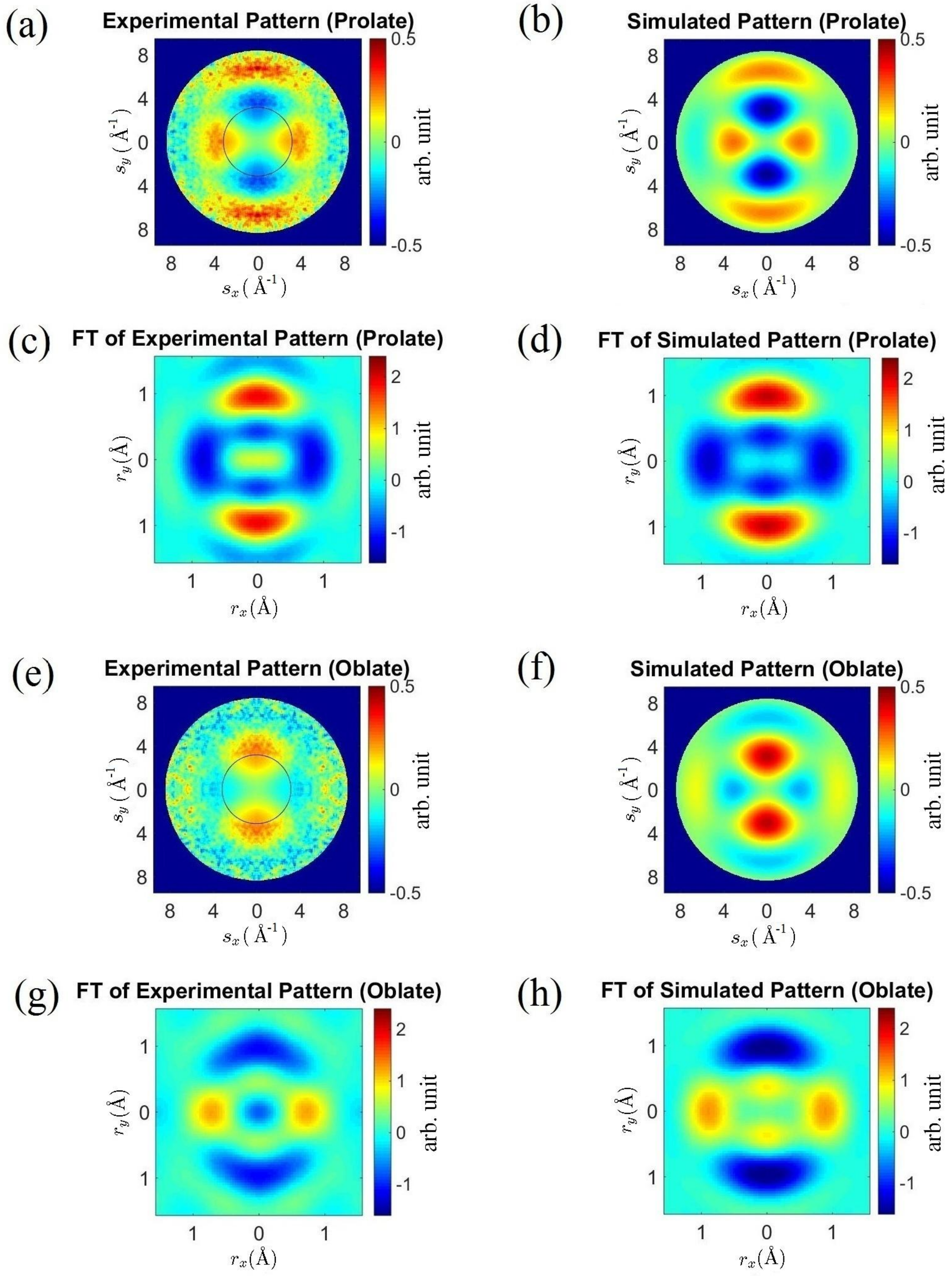

**Figure 4: 2-D $N_2$ diffraction patterns at half revival.**

**Spatial resolution of the molecular images.** The spatial resolution of the 2-D diffraction patterns shown in Fig. 4 can be determined in two different ways. First we can use equation (6), with $s_{max}$=8.3 Å$^{-1}$ we get the spatial resolution δ=0.76 Å. We can also determine spatial resolution directly from the autocorrelation images (Fig. 4c and 4g). For example, in Fig. 4c, the spatial resolution can be determined by converting the image into polar coordinates $o(r, \theta)$, then using a Gaussian function to fit to along the *r* dimension. The FWHM of the Gaussian fit was 0.76 Å using this method, consistent with the spatial resolution obtained using equation (5) and $s_{max}$ = 8.3 Å$^{-1}$.

**Angular Distribution based on 2-D images of aligned molecules.** The angular distribution of the molecules can be extracted from the patterns in Fig. 4c and 4g. The resulting distributions for prolate and oblate molecular ensembles are shown in Fig. 5a and 5b, respectively (see Angular Distribution section in Methods). The prolate distribution peaks at $\theta$ = 0 and 180 °, in the direction of the laser polarization. The oblate distribution peaks at $\theta$ = 90 °, in the direction perpendicular to the laser polarization. The degree of alignment is commonly measured by the quantity

$$\langle cos^2\alpha \rangle = \frac{\int_0^\pi f(\alpha) \cdot cos^2\alpha \cdot sin\alpha \cdot d\alpha}{\int_0^\pi f(\alpha) \cdot \sin\alpha \cdot d\alpha} \quad (6)$$

where $f(\alpha)$ is the angular distribution and $\alpha$ is the angle between the molecular axis and the laser polarization. A value of the alignment parameter $<\cos^2\alpha>=1$ corresponds to perfect alignment, whereas $<\cos^2\alpha>=1/3$ indicates random orientation. Any value between 1/3 and 1 indicates alignment, while any value below 1/3 indicates anti-alignment. In Fig. 5, three curves are displayed: $f(\alpha)$ extracted from data (red), $f(\alpha)$ obtained from simulation using the fluence and temperature extracted from the fitting routine (green), and the simulation results convolved with the 100fs RMS temporal resolution (blue). The temporal resolution is included by convolving a Gaussian pulse with the simulated temporal evolution of the angular distribution, and it has a significant effect because the distribution changes from prolate to oblate in 300 fs. The extracted angular distributions are in good agreement with the simulation for both prolate and oblate distributions, after taking into account the effect of the temporal resolution.

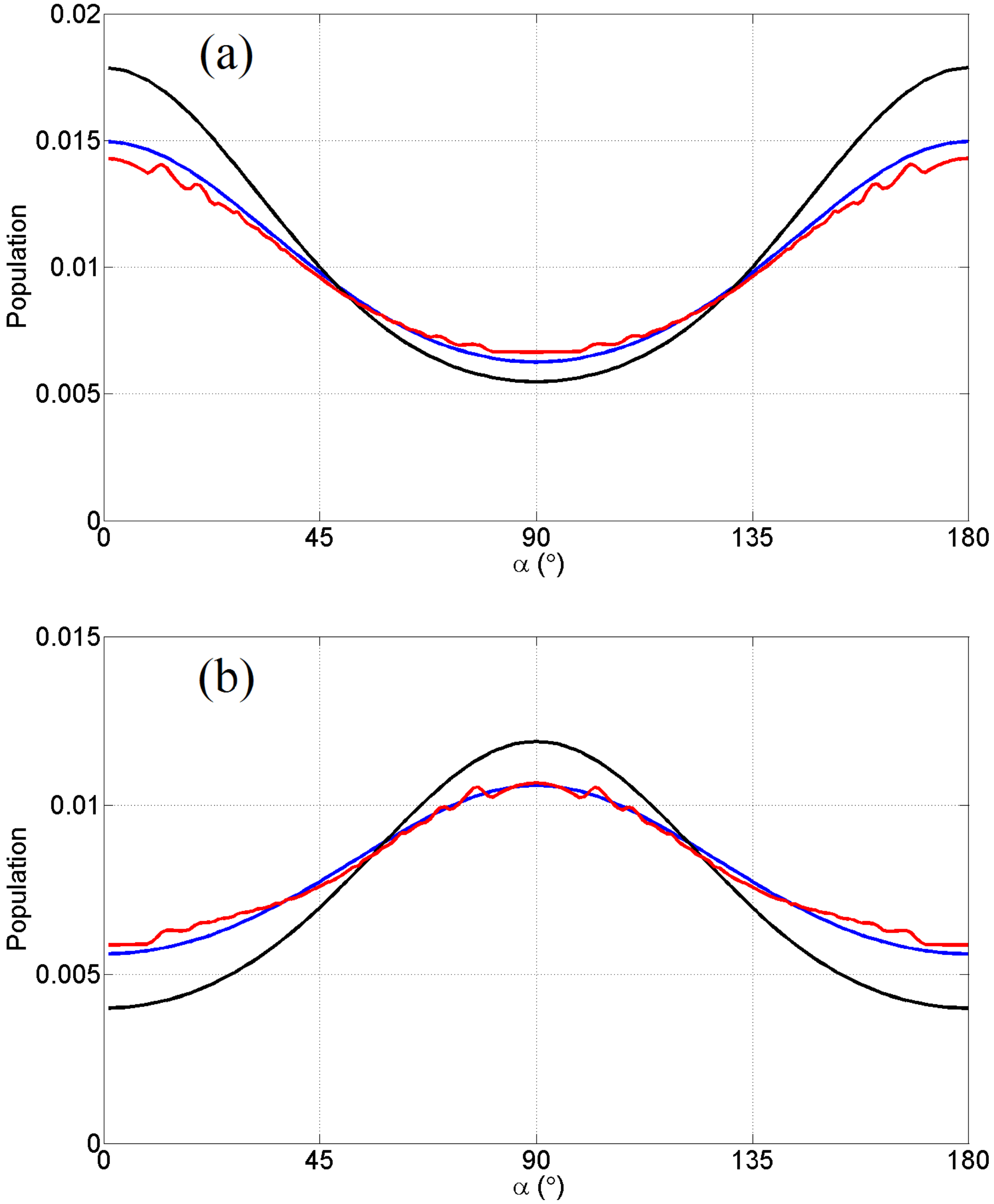

**Figure 5: Angular distributions.**

**Discussion**

In summary, we have shown that MeV UED can simultaneously reach a temporal resolution of 100 fs RMS and spatial resolution of 0.76 Å, which allows us to characterize the ultrafast evolution of a rotational wavepacket in $N_2$, measuring the angular distribution and the acquired molecular images with atomic resolution.

This achievement opens the door to a new class of experiments where changes in molecular geometry during a chemical reaction can be followed in space and time. In that context, the non-Born-Oppenheimer coupling induced by particular electronic potential energy topologies is of highest interest. The changing molecular geometry drives the molecule into those regions, where ultrafast intersystem crossing occurs, determining much of the molecular energy conversion in biology[4,27,28,32–36,66,67]. The current results demonstrate not only excellent temporal resolution, but also that sufficient signal-to-noise ratio for sub Angstrom spatial resolution can be acquired with MeV electrons. Nitrogen has a low atomic number, Z, and therefore a low scattering cross section, so we expect that the method will be successful for observing molecular dynamics in a large class of molecules.

For future developments, a number of upgrades can potentially improve both spatial and temporal resolution significantly. For example, a radio-frequency (RF) compression cavity could be used to compress the electron beam longitudinally[68], which can potentially lead to a temporal resolution on the order of 10fs together with orders-of-magnitude increase in charge per pulse. Time jitter and drift could be addressed by time-stamping techniques similar to those developed for XFELs[69]. CMOS active pixel sensor could potentially achieve single electron, single shot detection and thus eliminate detector noises[70].

## Methods

**Experimental Set-up.** The electron gun used in this experiment is a replica of the photo-injector used at the LCLS facility at SLAC National Accelerator Laboratory. A 15 fC per pulse charge is generated at the photocathode and 6 fC per pulse is delivered on target. A 50 nm thick $Si_3N_4$ membrane is used to separate the sample chamber and preserve the high vacuum in the electron gun. Roughly 60% of the charge is lost after the $Si_3N_4$ membrane and a 200 μm diameter collimator. The electron beam and pump laser are generated with a repetition rate of 120 Hz.

The $N_2$ gas is delivered into the chamber through a pulsed valve at a 120 Hz repetition rate. The backing pressure is 0.7 bar, and the nozzle orifice is 100 μm in diameter. During operation, the chamber pressure is approximately $6\times10^{-5}$ torr. The interaction region is roughly 350 μm away from the nozzle exit. The width of the laser beam, the electron beam and the gas jet are all around 200 μm FWHM at the interaction region. The sample-to-detector distance is calibrated with diffraction from a single crystal gold sample. There is a $\pm2.5\%$ uncertainty in the distance calibration, due to the quality of the gold sample and quality of the electron beam after the $Si_3N_4$ membrane.

Spatial alignment of the gas jet, the electron beam and the pump laser is obtained by positioning the focus of the laser approximately underneath the tip of the nozzle, and maximizing for the electron beam deflection by the plasma. Time zero can then be found, to within approximately 200 fs, by adjusting the delay between the pump and the probe beams and observing plasma deflection effects on the unscattered beam[23]. The laser focus is adjusted to 50 μm FWHM spot size for plasma lensing, giving a peak intensity of $8\times10^{14}$ W/cm$^2$. For the alignment experiments, the lens is driven 4 mm out of the focus to reach a spot size of 200 μm in the interaction region. The peak intensity for the experiment is $5\times10^{13}$ W/cm$^2$; only very faint plasma is observed in the interaction region for this intensity.

The detector comprises a phosphor screen, a ring-shaped mirror at a 45-degree angle to the beam, an f/0.85 lens and an electron multiply charge coupled device (EMCCD). A 4 mm diameter hole is drilled in the center of the phosphor screen to allow the unscattered electron beam to pass through. Data in the region $s < 3.5$ Å$^{-1}$ are not captured due to the hole in the detector.

**Data Processing of Diffraction Patterns and Anisotropy.** The experimental pattern is symmetrized over four quadrants, and the central region ($s < 3.5$ Å$^{-1}$) is extrapolated from existing data by letting the pattern smoothly go to zero towards the center.

The diffraction pattern of the prolate distribution is integrated over 90 minutes, and that of the oblate distribution is integrated over 60 minutes. The anisotropy is calculated by dividing the total counts in a horizontal cone by the total counts in a vertical cone for each diffraction pattern. The region used for this calculation is between $s = 3$ Å$^{-1}$ and $s = 4.5$ Å$^{-1}$. The horizontal cone has a half angle of 35° and the vertical cone has a half angle of 55°.

**Alignment Simulation.** The simulation of impulsive alignment is calculated using a linear rigid rotor interacting with a non-resonant pulse described by the time-dependent Schrodinger equation[62]. The temporal resolution is implemented by a convolution with a Gaussian function in time.

**Diffraction Pattern Simulation.** The simulations of diffraction patterns of a given angular distribution are calculated using an incoherent weighted sum of diffraction patterns from single molecules with different orientations. In the simulations, the atoms are assumed to be stationary at their equilibrium positions without vibrations[63].

**Temporal Evolution Fitting.** The temporal evolution fitting shown in figure 3a is obtained by a $\chi^2$ fitting of the simulation and data. The simulated anisotropy is obtained by an alignment simulation that gives angular distribution as a function of time, followed by a diffraction simulation according to the angular distribution. The anisotropy is calculated using the same method as for data, and the temporal resolution is obtained by a convolution of the calculated anisotropy with a Gaussian beam in time.

**Angular Distribution.** The angular distribution in Fig. 5a and 5b are extracted from Fig. 4c and 4g, respectively. The procedure includes three steps: (1) extracting the difference angular distribution from the 2-D difference pattern, (2) adding a baseline that accounts for the angular distribution of the reference pattern to the difference in angular distribution, and (3) normalizing the angular distribution. Step 1 is implemented by first converting Fig. 4c and 4g to polar coordinates, then integrating over the radial coordinate within the FWHM of the peak. In step 2, the baseline is obtained by simulation, corrected by the fitting parameter that accounts for the spatial overlap between laser and electrons (obtained from the fitting for Fig. 3a)). In step 3, the normalization used is $\int_0^\pi f(\alpha) \cdot \sin\alpha \cdot d\alpha$=1.

**References**

1. Pedersen, S., Herek, J. L. & Zewail, A. H. The validity of the 'diradical' hypothesis: direct femtoscond studies of the transition-state structures. *Science* **266,** 1359–1364 (1994).

2. Waldeck, D. H. Photoisomerization dynamics of stilbenes in polar solvents. *J. Mol. Liq.* **57,** 127–148 (1993).

3. Schoenlein, R. W., Peteanu, L. A., Mathies, R. A. & Shank, C. V. The first step in vision: femtosecond isomerization of rhodopsin. *Science* **254,** 412–415 (1991).

4. Mathies, R. A., Brito Cruz, C. H., Pollard, W. T. & Shank, C. V. Direct observation of the femtosecond excited-state cis-trans isomerization in bacteriorhodopsin. *Science* **240,** 777–779 (1988).

5. Weiss, S. Fluorescence Spectroscopy of Single Biomolecules. *Science* **283,** 1676–1683 (1999).

6. Milne, C. J., Penfold, T. J. & Chergui, M. Recent experimental and theoretical developments in time-resolved X-ray spectroscopies. *Coord. Chem. Rev.* **277,** 44–68 (2014).

7. Sil, S. & Umapathy, S. Raman spectroscopy explores molecular structural signatures of hidden materials in depth: Universal Multiple Angle Raman Spectroscopy. *Sci. Rep.* **4,** 5308 (2014).

8. Blanchet, V., Zgierski, M. Z., Seideman, T. & Stolow, A. Discerning vibronic molecular dynamics using time-resolved photoelectron spectroscopy. *Nature* **401,** 52–54 (1999).

9. Bisgaard, C. Z. *et al.* Time-resolved molecular frame dynamics of fixed-in-space CS2 molecules. *Science* **323,** 1464–1468 (2009).

10. Deb, S. & Weber, P. M. The ultrafast pathway of photon-induced electrocyclic ring-opening reactions: the case of 1,3-cyclohexadiene. *Annu. Rev. Phys. Chem.* **62,** 19–39 (2011).

11. Gessner, O. *et al.* Femtosecond multidimensional imaging of a molecular dissociation. *Science* **311,** 219–222 (2006).

12. Küpper, J. *et al.* X-ray diffraction from isolated and strongly aligned gas-phase molecules with a free-electron laser. *Phys. Rev. Lett.* **112,** 83002 (2014).

13. Ihee, H. *et al.* Ultrafast x-ray diffraction of transient molecular structures in solution. *Science* **309,** 1223–1227 (2005).

14. Siders, C. W. *et al.* Detection of nonthermal melting by ultrafast X-ray diffraction. *Science* **286,** 1340–1342 (1999).

15. Minitti, M. P. *et al.* Imaging Molecular Motion: Femtosecond X-Ray Scattering of an Electrocyclic Chemical Reaction. *Phys. Rev. Lett.* **114,** 255501 (2015).

16. Gao, M. *et al.* Mapping molecular motions leading to charge delocalization with ultrabright electrons. *Nature* **496,** 343–346 (2013).

17. Siwick, B. J., Dwyer, J. R., Jordan, R. E. & Miller, R. J. D. An atomic-level view of melting using femtosecond electron diffraction. *Science* **302,** 1382–1385 (2003).

18. Ihee, H. *et al.* Direct imaging of transient molecular structures with ultrafast diffraction. *Science* **291,** 458–462 (2001).

19. Srinivasan, R., Feenstra, J. S., Park, S. T., Xu, S. & Zewail, A. H. Dark structures in molecular radiationless transitions determined by ultrafast diffraction. *Science* **307,** 558–563 (2005).

20. Blaga, C. I. *et al.* Imaging ultrafast molecular dynamics with laser-induced electron diffraction. *Nature* **483,** 194–197 (2012).

21. Meckel, M. *et al.* Laser-induced electron tunneling and diffraction. *Science* **320,** 1478–1482 (2008).

22. Krasniqi, F. *et al.* Imaging molecules from within: Ultrafast angström-scale structure determination of molecules via photoelectron holography using free-electron lasers. *Phys. Rev. A - At. Mol. Opt. Phys.* **81,** 1–11 (2010).

23. Srinivasan, R., Lobastov, V. A., Ruan, C.-Y. & Zewail, A. H. Ultrafast electron diffraction (UED): A new development for the 4D determination of transient molecular structures. *Helv. Chim. Acta* **86,** 1763–1838 (2003).

24. Yang, J., Beck, J., Uiterwaal, C. J. & Centurion, M. Imaging of Alignment and Structural Changes of Carbon Disulfide Molecules using Ultrafast Electron Diffraction. *Nat. Commun.* accepted (2015).

25. Sciaini, G. & Miller, R. J. D. Femtosecond electron diffraction: heralding the era of atomically resolved dynamics. *Reports Prog. Phys.* **74,** 096101 (2011).

26. Miller, R. J. D. Mapping Atomic Motions with Ultrabright Electrons: The Chemists' Gedanken Experiment Enters the Lab Frame. *Annu. Rev. Phys. Chem.* 583–604 (2014). doi:10.1146/annurev-physchem-040412-110117

27. Levine, B. G. & Mart ́ınez, T. J. Isomerization Through Conical Intersections. *Annu. Rev. Phys. Chem.* **58,** 613–634 (2007).

28. Kochman, M. A., Tajti, A., Morrison, C. A. & Miller, R. J. D. Early Events in the Nonadiabatic Relaxation Dynamics of 4-( N , N -Dimethylamino)benzonitrile. *J. Chem. Theory Comput.* **11,** 1118–1128 (2015).

29. Ischenko, A. A., Ewbank, J. D. & Lothar, S. Structural kinetics by stroboscopic gas electron diffraction Part 1. Time-dependent molecular intensities of dissociative states. *J. Mol. Struct.* **320,** 147–158 (1994).

30. Elsayed-Ali, H. E. & Mourou, G. a. Picosecond reflection high-energy electron diffraction. *Appl. Phys. Lett.* **52,** 103–104 (1988).

31. Ruan, C. Y. *et al.* Ultrafast diffraction and structural dynamics: the nature of complex molecules far from equilibrium. *Proc. Natl. Acad. Sci. U. S. A.* **98,** 7117–7122 (2001).

32. Schultz, T. *et al.* Mechanism and Dynamics of Azobenzene Photoisomerization. *J. Am. Chem. Soc.* **125,** 8098–8099 (2003).

33. Crespo-Hernandez, C., Cohen, B., Hare, P. & Kohler, B. Ultrafast excited-state dynamics in nucleic acids. *Chem. Rev.* **104,** 1977–2019 (2004).

34. McFarland, B. K. *et al.* Ultrafast X-ray Auger probing of photoexcited molecular dynamics. *Nat. Commun.* **5,** 4235 (2014).

35. Middleton, C. T. *et al.* DNA Excited-State Dynamics: From Single Bases to the Double Helix. *Annu. Rev. Phys. Chem.* **60,** 217–239 (2009).

36. Schreier, W. J. *et al.* Thymine dimerization in DNA is an ultrafast photoreaction. *Science* **315,** 625–629 (2007).

37. Siwick, B. J., Dwyer, J. R., Jordan, R. E. & Miller, R. J. D. Ultrafast electron optics: Propagation dynamics of femtosecond electron packets. *J. Appl. Phys.* **92,** 1643–1648 (2002).

38. Dantus, M., Kim, S. B., Williamson, J. C. & Zewail, A. H. Ultrafast electron diffraction. 5. Experimental time resolution and applications. *J. Phys. Chem.* **98,** 2782–2796 (1994).

39. Wang, X., Qiu, X. & Ben-Zvi, I. Experimental observation of high-brightness microbunching in a photocathode rf electron gun. *Phys. Rev. E* **54,** R3121–R3124 (1996).

40. Wang, X. J., Wu, Z. & Ihee, H. Femto-seconds electron beam diffraction using photocathode RF gun. *Proc. 2003 Part. Accel. Conf.* **1,** 420–422 (2003).

41. Wang, X. J., Xiang, D., Kim, T. K. & Ihee, H. Potential of Femtosecond Electron Diffraction Using Near-Relativistic Electrons from a Photocathode RF Electron Gun. *J. Korean Phys. Soc.* **48,** 390–396 (2006)..

42. Hastings, J. B. *et al.* Ultrafast time-resolved electron diffraction with megavolt electron beams. *Appl. Phys. Lett.* **89,** 184109 (2006).

43. Musumeci, P., Moody, J. T. & Scoby, C. M. Relativistic electron diffraction at the UCLA Pegasus photoinjector laboratory. *Ultramicroscopy* **108,** 1450–1453 (2008).

44. Li, R. *et al.* Experimental demonstration of high quality MeV ultrafast electron diffraction. *Rev. Sci. Instrum.* **80,** 083303 (2009).

45. Muro'Oka, Y. *et al.* Transmission-electron diffraction by MeV electron pulses. *Appl. Phys. Lett.* **98,** 2009–2012 (2011).

46. Fu, F. *et al.* High quality single shot ultrafast MeV electron diffraction from a photocathode radio-frequency gun. *Rev. Sci. Instrum.* **85,** 083701 (2014).

47. Manz, S. *et al.* Mapping atomic motions with ultrabright electrons: towards fundamental limits in space-time resolution. *Faraday Discuss.* **177,** 467–491 (2015).

48. Zhu, P. *et al.* Femtosecond time-resolved MeV electron diffraction. *New J. Phys.* **17,** 063004 (2015).

49. Reiser, M. *Theory and Design of Charged Particle Beams*. (New York: Wiley, 1994).

50. Stapelfeldt, H. & Seideman, T. Colloquium: Aligning molecules with strong laser pulses. *Rev. Mod. Phys.* **75,** 543–557 (2003).

51. Rosca-Pruna, F. & Vrakking, M. J. Experimental observation of revival structures in picosecond laser-induced alignment of I2. *Phys. Rev. Lett.* **87,** 153902 (2001).

52. Hensley, C. J., Yang, J. & Centurion, M. Imaging of isolated molecules with ultrafast electron pulses. *Phys. Rev. Lett.* **109,** 133202 (2012).

53. Reckenthaeler, P. *et al.* Time-resolved electron diffraction from selectively aligned molecules. *Phys. Rev. Lett.* **102,** 213001 (2009).

54. Chen, Y.-H., Varma, S. & Milchberg, H. M. Space- and time-resolved measurement of rotational wave packet revivals of linear gas molecules using single-shot supercontinuum spectral interferometry. *J. Opt. Soc. Am. B* **25,** B122 (2008).

55. Litvinyuk, I. V *et al.* Alignment-dependent strong field ionization of molecules. *Phys. Rev. Lett.* **90,** 233003 (2003).

56. Itatani, J. *et al.* Controlling high harmonic generation with molecular wave packets. *Phys. Rev. Lett.* **94,** (2005).

57. McFarland, B. K., Farrell, J. P., Bucksbaum, P. H. & Gühr, M. High harmonic generation from multiple orbitals in N2. *Science* **322,** 1232–1235 (2008).

58. Cryan, J. P. *et al.* Auger electron angular distribution of double core-hole states in the molecular reference frame. *Phys. Rev. Lett.* **105,** (2010).

59. Weathersby, S. P. *et al.* MeV Ultrafast Electron Diffraction at SLAC. *Rev. Sci. Instrum.* **86**, 073702 (2015).

60. Salvat, F., Jablonski, A. & Powell, C. J. Elsepa - Dirac partial-wave calculation of elastic scattering of electrons and positrons by atoms, positive ions and molecules. *Comput. Phys. Commun.* **165,** 157–190 (2005).

61. Huber, K. P. & Herzberg, G. in *Molecular Spectra and Molecular Structure* (1979). doi:10.1007/978-1-4757-0961-2_2

62. Ortigoso, J., Rodriguez, M., Gupta, M. & Friedrich, B. Time evolution of pendular states created by the interaction of molecular polarizability with a pulsed nonresonant laser field. *J. Chem. Phys.* **110,** 3870–3875 (1999).

63. Yang, J., Makhija, V., Kumarappan, V. & Centurion, M. Reconstruction of three-dimensional molecular structure from diffraction of laser-aligned molecules. *Struct. Dyn.* **1,** 044101 (2014).

64. Holmegaard, L. *et al.* Control of rotational wave-packet dynamics in asymmetric top molecules. *Phys. Rev. A* **75,** 051403 (2007).

65. Hagena, O. F. Nucleation and growth of clusters in expanding nozzle flows. *Surface Science Letters* **106,** 101–116 (1981).

66. Schenkl, S., van Mourik, F., van der Zwan, G., Haacke, S. & Chergui, M. Probing the ultrafast charge translocation of photoexcited retinal in bacteriorhodopsin. *Science* **309,** 917–920 (2005).

67. Polli, D. *et al.* Conical intersection dynamics of the primary photoisomerization event in vision. *Nature* **467,** 440–443 (2010).

68. Li, R. K., Musumeci, P., Bender, H. a., Wilcox, N. S. & Wu, M. Imaging single electrons to enable the generation of ultrashort beams for single-shot femtosecond relativistic electron diffraction. *J. Appl. Phys.* **110,** 1–10 (2011).

69. Beye, M. *et al.* X-ray pulse preserving single-shot optical cross-correlation method for improved experimental temporal resolution. *Appl. Phys. Lett.* **100,** 1–5 (2012).

70. Battaglia, M. *et al.* Characterisation of a CMOS active pixel sensor for use in the TEAM microscope. *Nucl. Instruments Methods Phys. Res. Sect. A Accel. Spectrometers, Detect. Assoc. Equip.* **622,** 669–677 (2010).

**List of Figures**

**Figure 1: Experimental layout.** A sketch of the experimental setup. A 3.7 MeV pulsed electron beam (blue) is directed towards a nitrogen gas jet (grey). The gas jet is introduced into the vacuum chamber using a pulsed nozzle (black). The pump laser pulse (red) is deflected by two ring-shaped mirrors. The laser propagates at a small angle (~5 °) with respect to the electron beam. The diffraction pattern is recorded with a phosphor screen located 3.1m downstream from the interaction region. The unscattered electron beam is transmitted through a hole in the phosphor screen.

**Figure 2: Static $N_2$ diffraction.** Theoretical (blue) and experimental (red) modified diffraction intensity *sM* from $N_2$ gas.

**Figure 3: Temporal evolution of the $N_2$ rotational wavepacket.** Anisotropy in the diffraction patterns from experimental data (red) and simulation (blue) vs time. The simulated prolate (alignment) and oblate (anti-alignment) angular distributions are shown as insets at the half revival around 4 ps. The simulation parameters of initial rotational temperature, alignment laser fluence, temporal resolution and rescaling factor are obtained from a fitting routine. Each data point is accumulated over 2 minutes. $\chi^2$ error vs RMS temporal resolution in the four-parameter fit is shown in the inset.

**Figure 4: 2-D $N_2$ diffraction patterns at half revival.** (a) experimentally measured and (b) simulated diffraction-difference patterns of the prolate distribution. (c) and (d) are Fourier transforms of (a) and (b), respectively. (e) experimentally measured and (f) simulated diffraction-difference pattern of the oblate distribution. (g) and(h) are Fourier transforms of (e) and (f), respectively. In patterns (a) and (e), the data inside the black circles are missing due to the beam

stop. They are obtained by extrapolating the pattern and letting the counts smoothly go to zero towards the center.

**Figure 5: Angular distributions.** (a) Prolate angular distribution at the revival: Experimental (red), simulated (black) and simulated convolved with 100 fs RMS temporal resolution (blue). The $\langle \cos^2\alpha \rangle$ values for red, black and blue curve are 0.41, 0.45 and 0.42, respectively. (b) Oblate angular distribution: Experimental (red), simulated (black) and simulated convolved with 100 fs RMS temporal resolution (blue). The $\langle \cos^2\alpha \rangle$ values for red, black and blue curve are 0.28, 0.25 and 0.28, respectively. The experimental curves are measured from Fig. 4c and 4g, respectively.

**Acknowledgements**

The authors would like to thank SLAC management for the strong support. The technical support by SLAC Accelerator Directorate, Technology Innovation Directorate, LCLS Laser Science & Technology division and Test Facilities Department is gratefully acknowledged. This work was supported in part by the U.S. Department of Energy (DOE) Contract No. DE-AC02-76SF00515, DOE Office of Basic Energy Sciences Scientific User Facilities Division, and the SLAC UED/UEM Initiative Program Development Fund. J. Yang, and M. Centurion were partially supported by the U.S. Department of Energy Office of Science, Office of Basic Energy Sciences under Award Number DE-SC0003931. M. S. Robinson was supported by the National Science Foundation EPSCoR RII Track-2 CA Award No. IIA-1430519.

## Author contributions

J.Y., M.G., T.V., M. S. R., R.L., X.S., T. G., F. W., S.W, and X.W. carried out the experiments. N.H., R. C., J. C., I. M., S. V., and A.F. developed the laser system. M.G. and J.Y. constructed the setup for gas phase experiments. C. H., K. J., A. R., and C. Y. helped on experimental setup. J.Y. performed the data analysis and simulations. The experiment was conceived by M.G., M.C. and X.W. The manuscript was prepared by J.Y., M.C., M. S. R., and M.G with discussion and improvements from all authors. M.C. and X.W. supervised the work.